# Irving Segal's Axiomatization of Spacetime and its Cosmological Consequences[1]:

## The Cosmological Redshift in the Einstein Static Universe And Energy Conservation

Aubert Daigneault, professeur honoraire,
Département de mathématiques et de statistique, Université de Montréal.
Email: aubert.daigneault@umontreal.ca

---

[1] Invited lecture given at the conference celebrating the 100th anniversary of the Hungarian mathematical logicians László Kalmár and Rózsa Péter in Budapest, August 5 – August 11, 2005. Dedicated to the memory of my mother, Lorette Sauriol Daigneault, who would also be 100 years old in 2005.

# 1 Abstract

Irving Ezra Segal (1918-1998), an MIT mathematician, has devoted much of his life to an axiomatic theory of spacetime, called chronometric cosmology (CC), which is generally ignored by astrophysicists. The axioms are properties of Minkowski spacetime $M_0$ and admit only one other model $M$ which can briefly be described as the supposedly discredited cosmological model known as the Einstein universe first proposed by Einstein in 1917 [Ein17]. CC assumes special relativity (SR) as a local theory inasmuch as this can be identified with $M_0$. Otherwise CC does not assume general relativity (GR) but is compatible with it. Hence Segal's approach to $M$ is quite unlike that of Einstein.

CC attributes a central role to the causal structure of spacetime which represents the relation of temporal precedence. Spacetime defined as the totality of all events - past, present and future - is, before anything else, a partially ordered set: the relation $p = q$ between two distinct events $p$ and $q$ means that $p$ precedes $q$ i.e. the event $p$ occurred before the event $q$ (for instance, my mother's birth absolutely preceded mine). This relation, known as the relation of *causality*, of *temporal precedence*, or of *anteriority*, is the most immediate observational data. It is conceptually and psychologically more fundamental than the measurement of distances and durations, and is independent of any observer.

All but one of Segal's axioms are verified in many models of spacetime. The most controversial assumption is that of *causal temporal homogeneity* which is responsible for the principle of conservation of energy. Fundamental as it is in science generally, in particular in classical and quantum mechanics and in SR, this principle holds only in the form of a differential conservation of energy-momentum in GR. It holds naturally in the two models $M_0$ and $M$. The model $M$ has been abandoned by the astronomical community as a result of the discovery of the cosmological redshift phenomenon, which, interpreted as expansion of the universe, appears to be incompatible with it, $M$ being eternal and static i.e. non expansionary.

Contrary to this view, the redshift phenomenon, is a property of $M$ which appears as a theorem of CC. The proof rests on an analysis of the geometrical relationship between $M_0$ and $M$ that can be visualized as the well known one between the complex plane and the Riemann sphere. This leads to a CC redshift formula quite different from the Hubble law. Segal has relentlessly analysed statistically the two competing laws with all astronomical data available up to his untimely death. His contention is that the CC formula is phenomenologicaly tenable whereas the widely believed Hubble law is not.

The fundamentals of CC can be found in [Seg76.1] and in [Lev95]. All of Segal's books and papers are listed in [Gro02].

I am most thankful to Alexander Levichev for the many improvements he has suggested to a preliminary version of this paper.

## 2   Introduction

The axioms *of chronometric cosmology* (CC) as Irving Segal called his cosmology can be summed up as a sort of global cosmological principle asserting the global *causal* isotropy and homogeneity of both, space and time.

Since they refer only to a causal structure and assume no metric on spacetime, contrary to what is usually the case, some appear as watered down or conservative versions of the familiar ones embodied in the definition of Friedmann-Lemaître-Robertson-Walker (FLRW) spacetimes. On the other hand, the ones referring to the isotropy and homogeneity of time are more demanding. All, however, are properties of Minkowski spacetime. The net effect of theses axioms is to yield a larger symmetry group of spacetime and thus reduce the number of models to just two: Minkowski spacetime $M_0$ and the Einstein static universe $M$.

As just said, he axioms of CC make no reference to any metric on spacetime. However, one may always assume that such a metric is present by virtue of independent work of E. B. Vinberg and of Jacques Tits. One must be careful to distinguish between causal mappings and metric preserving maps. For instance, an isometry is a causal automorphism but not inversely. Besides, the metric need not be unique.

Also contrary to what is generally assumed the axioms have nothing to do with any uniformity in the distribution of matter in space or time as, in CC, the global shape of space, its origin and destiny are independent of its material and energetic content.

The cosmological consequences of these assumptions fly in the face of present day dogmas in cosmology: the universe is eternal; there is no such thing as the expansion of the universe and no such thing as a big bang; space is a hypersphere i.e. a three-dimensional sphere of fixed radius; the principle of energy conservation is reestablished; the redshift phenomenon is no Doppler effect but is an effect of the curvature of space, a possibility imagined by Hubble and Tolman in 1935 [H/T35].

There is a twofold geometrical connection between the two models. On the one hand $M_0$ is causally embedded into $M$ by a relativistic generalization of stereographic projection, and on the other hand it is tangent to $M$ at the point of observation, as is the case for any Lorentzian (also called hyperbolic) manifold.

In this lecture we will not demonstrate the nonexistence of other models but, rather, will, after stating the axioms, concentrate on a detailed description of the geometrical relationship between $M_0$ and $M$ and its main cosmological consequences: the mechanism of the redshift in $M$ and energy conservation.

Here are the main steps in establishing this relationship:

First the role of Minkowski spacetime will be played by the causally isomorphic real linear space *H(2)* of Hermitian matrices of order 2 and that of the three-sphere $S^3$ by the isometric multiplicative group *SU(2)* of special unitary matrices of order 2.

Second *H(2)* will be causally immersed  into the group *U(2)* of all unitary matrices of order 2.  This immersion called the *Cayley transform*, turns out to be the inverse of a generalized stereographic projection.

Finally, the conformal compactification *U(2)* of *H(2)* will be lifted to its universal covering space which turns out to be *M*.  Thus a sequence of mappings is constructed the net effect of which is to causally immerse Minkowski spacetime $M_0$ into the Einstein universe *M*

$$M_0 \rightarrow H(2) \rightarrow U(2) \rightarrow R^1 \times SU(2) \rightarrow R^1 \times S^3 = M \qquad (1)$$

The generalized stereographic projection parallels the one between the unit circle in the complex plane, a multiplicative Lie group,  and the imaginary axis its Lie algebra. The linear space *H(2)* can be viewed as the Lie algebra of the Lie group *U(2)*.  (This could be similarly generalized, for all *n*, to *H(n)* and *U(n)*).

The lifting of *U(2)* into its universal covering is to avoid  time being circular in accordance with one of the postulates.

Time differs crucially in the two models and we will therefore be speaking of Minkowskian or local time $x_0$ and of Einsteinian or cosmic time *t*.  The relation between the two times is non-linear and is given by the following *two-times formula*, which we will demonstrate, and in which *r* stands for the radius of the three-sphere in the model *M*

$$x_0 = 2r \tan(t/2r) \qquad (2)$$

We assume throughout that appropriate units are used such that the speed of light *c* is 1.

In particular, this implies that the whole infinite local time line in $M_0$ corresponds to a finite interval, from $-\pi r$ to $+\pi r$, on the infinite cosmic time line in *M*.  This means that, according to CC, what we humans perceive as eternity in both directions is only a finite portion of real eternal time in both directions.

Which time coordinate is actually measured in local observations of physical phenomena of relatively small spacetime extent is empirically immaterial, as the two differ by unobservably small amounts. Indeed, assuming

> *"as is commonly believed that $r \geq 10^8$ light-years", the two "deviate by less than one part in $10^{15}$ out to distances of 1 light-year, or of less than 1 part in $10^6$ out to galactic distances.  There is no apparent means to detect such differences in classical observations." [Seg74, p. 765]*

The same question arises with space coordinates.

Indeed an application of l'Hospital's rule to equation (2) shows that $x_0$ tends to $t$ as $r$ tends to infinity as expected. The above mentioned unobservably small differences between the two times can easily be established on the basis of the series expansion of $x_0$ in powers of $t$, which starts as follows:

$$x_0 = t + t^3/(12r^2) + t^5/(120r^4) + ... \qquad (3)$$

As a result, only extragalactic observations can be relevant for telling the two times apart. As Segal writes:

> "...in the absence of precise observations of masses and distances outside the solar system, only the redshift and related observations appear to fall in the category required to distinguish, potentially at least, between the two clocks." [Seg82, p. 857]

Concurrently with the two-times formula (2) the *redshift formula*

$$z = \tan^2(t/2r) \qquad (4)$$

giving the observed redshift $z$ after a time of propagation $t$, or equivalently, a geodesic distance on the 3-sphere, can be derived.

Equation (4) reveals that for small values of $t$ (or, equivalently, of the distance), the redshift varies as the square of $t$ in contradiction with Hubble's law which is linear. From (4), we also see that as $r$ tends to infinity, $z$ tends to 0 which means that as the curvature of space disappears, the redshift follows suit.

In recent years the much heralded, and for long unexpected claim, of an acceleration in the rate of universal expansion of the universe has retained attention. The claim is based on the observation that the rise and fall of the light intensity of distant supernovae is over a longer time span than similar events in nearby supernovae. It is observed that this time span is stretched by a factor of $1 + z$. Chronometric cosmology offers a simple explanation of this dilation effect simply by differentiating the two-times formula (2) and obtaining

$$\frac{\partial x_0}{\partial t} = \sec^2(t/2r) = 1 + \tan^2(t/2r) = 1 + z \qquad (5)$$

To these two times are related two different concepts of energy: the Einsteinian or cosmic and the Minkowskian or local with the former exceeding the latter.

In CC the redshift represents the difference between the Einstein energy and the Minkowski energy. The well known formula $h\nu$ where $h$ is the Planck constant and $\nu$ the frequency of a photon wave which is held to be the energy of that photon is, according to CC, only the Minkowskian energy and is valid only for recently emitted photons. The true cosmic energy is larger; the difference increases with $z$ and represents the apparent loss of energy attributed to the lower frequency i.e. the redshift.

The aim of the lecture is to state the axioms; describe in detail the relationship between the two models; establish the two-times formula and the redshift formula; and establish energy conservation in the two models.

We will not discuss here the phenomenological side of the issue. We have devoted a paper [Dai03] to that and many references can be found in the bibliography prepared by Leonard Gross [Gro02]. Let it be said though, that, as a whole, the astronomical community, a priori convinced of the expansion of the universe, has shown little interest in –and even anger at- the numerous statistical studies which systematically point to the contrary.

No formalization of the axioms of CC or of derivations based on them in the mathematical logicians' sense; in particular, in first order logic, is attempted. However the theory is axiomatic and certainly belongs to the foundations of spacetime. Maybe formalization could be pursued. We leave that to the Budapest logicians.

## 3  Statement of the axioms

The axioms of CC can be summed up as the following general conditions embodying three fundamental physical principles: the *isotropy* and *homogeneity* of space, i.e. the absence of a preferred direction at any point of space and the absence of a preferred point in space; second, the "*principle of inertia*", i.e. the statement that there is no preferred timelike direction, which means the equivalence between observers in relative motion at the same point; and third, the possibility to *globally* factor spacetime as *time × space* and *temporal homogeneity* with respect to this factorization.

This requirement of temporal homogeneity is the main controversial axiom.

Segal opined that such principles should be abandoned only in the face of compelling evidence to the contrary. He contended that not a shred of such evidence exists and that, on the contrary, the verifiable consequences of these principles are confirmed by all available astronomical data.

Speaking apparently of at least the first three axioms to be stated, Segal wrote: "The following axioms seem about as primitive and unexceptionable as, say, the axioms of Peano for the integers." [Seg80, p. 386]

### 3.1   Axiom 1: Spacetime is a differential manifold of dimension four

### 3.2   Axiom 2: Spacetime is endowed with a causal structure.

The concept of causal structure axiomatizes what one obtains in any Lorentzian manifold by discarding the metric and retaining only the assignment to each point of the manifold of the closed cone of future directions in the tangent space at that point.

To be precise we need definitions.

A convex cone in a real linear vector space is a set $C$ of vectors, containing a nonzero vector and which is closed under addition and under multiplication by non negative constants.

The cone $C$ is non trivial if $C$ and $-C$ have only 0 in common.

The axiom means that there is given a distinguished non trivial closed convex cone field on spacetime i.e. a smooth assignment $p$ a $C(p)$ to each point $p$ of spacetime of a non trivial closed cone $C(p)$, called the *future cone* at $p$ (and representing physically the totality of infinitesimal future directions as perceived by an observer at $p$), in the tangent space at $p$ such that the associated relation of *temporal precedence* to be defined in a moment is transitive.

The non triviality is meant to exclude Newtonian causality and thus insinuates Einstein's requirement that there is a finite limit velocity to light and other physical processes. The closure property is meant to include in the cone both *timelike* and *lightlike* directions. A direction is said to be timelike if it is in the interior of *C(p)* and lightlike if it is on its boundary [Seg 80, p. 386], [Seg76.1, p. 22, p. 25].

By definition an event $p$ *temporally precedes* a distinct event $q$, and we write $p = q$, if there exists an oriented curve going from $p$ to $q$ at every point of which the forward pointing tangent belongs to the future cone at that point. Such a curve is called a timelike curve or arc; it is called lightlike if at each point of the arc, the tangent belongs to the boundary of the cone at that point [Seg76.1, p. 27].

### 3.3   Axiom 3: Time does not wind back on itself

This means that the causal structure of spacetime is causally global in the sense that it admits no timelike loops i.e. no closed timelike curves. In other words this means that if $p$ temporally precedes $q$ then $q$ does not temporally precede $p$ [Seg76.1, p.51], [Seg80, p. 387].

> *"[That] time would "wind back" on itself in the long run ...is counter to intuition, thermodynamics, and general physical ideas; while certain microscopic physical phenomena may well be cyclical in time, cyclicity of* *the Cosmos as a whole is generally implausible. We shall assume -without prejudice to future possibilities – that this is not the case."*
*[Seg76.1, p. 52]*

### 3.4  Axiom 4: The principle of causal spatial isotropy

This means that at any point of spacetime there is no preferred spacelike direction i.e. given two spacelike directions at any point *p* of spacetime, i.e. two directions not in *C(p) nor in –C(p),* there exists a causal diffeomorphism of spacetime onto itself that maps one of these directions on the other [Seg76.1, pp. 27, 55, 56].

Again, this is not meant to imply spatial uniformity in the distribution of matter; Euclidean geometry as normally used  does not imply that either.

### 3.5  Axiom 5: The principle of causal temporal isotropy

This is the statement of the equivalence between observers in relative motion at the same point. [Seg80, p. 393]

This means that there is no preferred timelike direction at any point of spacetime. More specifically, given two timelike directions at a given point of  spacetime, there is a causal diffeomorphism of spacetime onto itself that maps one of these directions on the other. Put otherwise, the group of causal diffeomorphisms of spacetime onto itself acts transitively on the timelike directions at any point of spacetime.

In the case of a Lorentzian manifold inducing the causal structure, one may be tempted to require further that this causal diffeomorphism be an isometry.  Such a stronger requirement is indeed valid in de Sitter spacetime [Tits79, p. 172-173] but it is not valid in Einstein spacetime for instance [Seg76.1, p. 59].

As we are celebrating 1905, it is proper to recall (in translation) the first statement of this *principle of relativity* as it appeared  in one of Einstein 1905 papers:[Ein05, p. 69 in the translation]:

> *"The laws by which the states of physical systems alter are independent of the alternative, to which of two systems of coordinates, in uniform motion of parallel translation relatively to each other, these alterations of state are referred (principle of relativity)."*

### 3.6  Axiom 6 : Spacetime can be globally factorized into time and space

Every event takes place some time, somewhere.

This means that there exists a three dimensional manifold $S$ called *space* and a diffeomorphism $\phi$ of the direct product of the *real time line R* with $S$ onto spacetime such that:

- for any $x$ in $S$ the map $t \rightarrow \phi(t, x)$ is a timelike arc in spacetime; and
- for any fixed $t$ in $R$, the map $x \rightarrow \phi(t, x)$ defines a spacelike submanifold of spacetime.

A spacelike submanifold is defined as one in which any two points are unrelated by the relation of temporal precedence [Seg76.1, p. 27, 44].

### 3.7   Axiom 7:  The principle of causal temporal homogeneity.

This is the main controversial axiom.  It is needed for a *natural* principle of energy conservation i.e. one asserting that the energy is invariant under a group of causal temporal translations related to a factorization of spacetime as *time × space*.   Segal held it to be "*philosophically reasonable but physically tentative*" [Seg80, p. 390, 391], [Seg76.1, p. 58].

This is to be understood as requiring  not only the absence of a preferred moment on the time axis in the factorization of spacetime as *time × space* whose existence is asserted in Axiom 6, but also that time translations with respect to this factorization make up a group of *causal* automorphisms of spacetime, the *temporal group* belonging to this factorization.

This group is a one-parameter subgroup $\{T_t \,|\, t \in R\}$ of the causal group such that the time translation $T_t$ carries the point *(t', x)* of spacetime into *(t'+t, x)*.

The key word in Axiom 7 is *causal*.  A one-parameter group of time translations i.e. of timelike diffeomorphisms  by which we mean that the tangent vector to the trajectory of any point is timelike at any of its points, can be defined likewise in any Friedmann-Lemaître-Robertson-Walker (FLRW) spacetime whose time axis is the whole real line, such as de Sitter spacetime, but this group is not causal in general [R/C03].

### 3.8   First corollary : spatial homogeneity

In CC homogeneity of space means that for any two points of space in Axiom 6, there exists a causal diffeomorphism of spacetime which maps one of these points on the other and which preserves this factorization [Seg76.1, p. 56, 57].  More specifically, this means that given any two points $x$ and $x'$ of space $S$, there exists a causal diffeomorphism which maps *(t, x )* on *(t, x')* for all time $t$.

Homogeneity of space is generally presented as a consequence of isotropy [H/E73 p. 135], [M/T/W73, p.723], [Rin77, p. 202]. Together with isotropy of space, this property of spacetime is called the *cosmological principle*.

### 3.9   *Second corollary: causal isotropy of light rays*

More specifically, given two *lightlike* directions at a given point of spacetime, there is a causal diffeomorphism of spacetime that maps one of these directions on the other. Put otherwise, the group of causal diffeomorphisms of spacetime acts transitively on the *lightlike* directions at any point of spacetime. This follows easily from the spacelike causal isotropy but not conversely. [Tits60, pp. 107, 109, 112, 113], [Tits56, pp.46, 47]

We note that both de Sitter spacetime and Einstein spacetime satisfy the stronger condition in which the causal diffeomorphism is an isometry [Tits60, p. 110, 111].

## 4   The only two models of the axioms

### 4.1   *Minkowski spacetime and the Einstein universe*

Segal asserts that there are only two models of these axioms; namely, Minkowski spacetime and the Einstein universe [Seg76.1, p. 58], [Seg80, p393-394], [Seg90, p. 167]. In any case, no other model is known. We will not offer here a proof of the nonexistence of other models.

Segal writes [Seg80, p. 393]:

> "*The work of Vinberg classifying homogeneous cones limits the causal cones C(p) to cones definable in each tangent space by a quadratic equation, and thus implies that the causal structures must be induced from a Lorentzian metric. The work of Tits classifying Lorentzian manifolds enjoying various isotropy features shows that the cosmos must be locally Minkowskian, and in fact one of an explicitly enumerated set of possibilities.*" [Vin63], [Tits60, pp 109, 112, 113, 117-119], [Seg84, pp. 250, 252]

Minkowski spacetime $M_0 = R \times R^3$ has the usual hyperbolic metric $(dx_0)^2 - (dx_1)^2 - (dx_2)^2 - (dx_3)^2$.

The Lorentzian metric on the Einstein universe $M = R \times S^3$ is $c^2 dt^2 - r^2 ds^2$ where $dt$ is the ordinary metric on the real time line $R$, $c$ is the speed of light and $rds$ is the ordinary metric on a sphere $S^3$ of radius $r$. There is an eight parameter family of factorizations of *M,* which correspond biuniquely to such metrics. This is quite unlike what is the case in Minkowski spacetime.

### 4.2   *De Sitter spacetime is not a model*

Though de Sitter spacetime is diffeomorphic to the product of the real line with a three-sphere it is not *globally* factorizable as *time × space* in the sense of Axioms 6 and 7 [Wolf84, p. 68, (2.4.6)], [Seg76.1, p. 53, 58, 6], [P/R86, p. 337].

This is in spite of the fact that the isometry group of de Sitter spacetime is of dimension 10 (which is the dimension of the Poincaré group and hence the maximal number compatible with general relativity) enabling that many conservation laws in the ordinary sense [R/C03].

This spacetime has symmetries i.e. isometries not present in the Einstein universe whose group of isometries is of dimension 7. For instance, unlike what is the case in the Einstein universe, there is always an *isometry* mapping a given spacelike direction on another given one at any point of de Sitter spacetime [Tits60, p. 109, 112]. The same is true of timelike directions [Tits79, p. 173], [Seg76.1, top of p. 59].

There may well exist temporal isometry (and hence causal) groups in de Sitter spacetime but they are not properly related to a global factorization of that spacetime as *time × space*. The group of temporal translations in a FLRW space, in particular in de Sitter spacetime, is not an isometry group and it is not even causal.

We are here dealing with the so-called *spherical* de Sitter spacetime, which is the universal cover of the *elliptic* one [Sch56, p. 7-14], [Tits60, (3) p. 109; (4) p. 110, p. 114].

This (spherical) de Sitter spacetime is defined as the following affine quadric in 5 dimensional space [H/E73, p. 124], [Rind77, p. 185, (8.158)], [Tits60, p. 110]:

$$x_1^2 - x_2^2 - x_3^2 - x_4^2 - x_5^2 = -a^2$$

in which *a* is a non zero constant.

The Lorentzian metric of the quadric is induced by the following metric of the surrounding 5-dimensional space

$$b_1^5(x, y) = x_1 y_1 - \sum_{j=2}^{5} x_j y_j$$

This quadric is a pseudosphere [Wolf, p. 67, (2.4.5a)].

# 5 Causality relation in Minkowski spacetime $M_0$

Let us be quite explicit about the causal structure on Minkowski spacetime $M_0$.

Let Q denote the quadratic form of $M_0$ i.e. $Q(x) = x_0^2 - x_1^2 - x_2^2 - x_3^2$ where $x = (x_0, x_1, x_2, x_3)$ is an arbitrary vector in $M_0$.

Let $x = (x_0, x_1, x_2, x_3)$ and $y = (y_0, y_1, y_2, y_3)$ be elements of $M_0$. We define $x = y$ to mean that $y - x$ is a timelike vector oriented towards the future i.e. Q(y-x) > 0 and $x_0 < y_0$. A one-to-one mapping $F$ of $M_0$ onto itself is called a causal automorphism iff both $F$ and its inverse preserve this partial ordering. One can define a different relation $x < y$ by requiring that $y - x$ be lightlike i.e. null instead of timelike. It turns out [Nab92, p. 65] that $F$ is a causal automorphism iff both $F$ and its inverse preserve this other relation. The *light cone* or *null cone* at the origin is defined as $C_N(O) = \{x \mid Q(x) = 0\}$

The *future light cone* or *future null cone* at the origin is defined as $C_N^+(O) = \{x \mid Q(x) = 0; x_0 \geq 0\}$.

The *future time cone* at the origin is defined as $C_T^+(O) = \{x \mid Q(x) > 0; x_0 > 0\}$. It is the 'interior' of the null cone.

The *closed future cone* is the union of the future light cone and the future time cone.

A vector x is *timelike* if Q(x) > 0. It is *spacelike* if Q(x) < 0. It is *lightlike* if Q(x) = 0.

The cones at points other than the origin are obtained by translations from the ones at the origin.

## 5.1 The Alexandrov-Zeeman theorem

The relation of anteriority is well known in Minkowski spacetime and determines its ordinary Lorentzian metric to within a constant strictly positive factor. This latter fact is a nontrivial but fundamental theorem established in 1953 by the Russian mathematicians A. D. Alexandrov and V. V. Ovchinnikova [A/O53] and rediscovered a decade later by E. C. Zeeman [Zee64]. It entails that causality preserving maps between Lorentzian manifolds are the same as conformal maps preserving time orientation [Nab92, pp. 64-74], [S/W95, p. 310, 311]. A Lorentz transformation is said to be orthochronous if it preserves time orientation. The theorem can be stated thus:

> **Theorem**: The group of causal automorphisms of $M_0$ is the orthochronous Poincaré group also called the inhomogeneous orthochronous Lorentz group. Any element $F$ of this group can be written as $F = T \circ K \circ L$ where $L$ is an orthochronous Lorentz transformation , $K$ is a dilation and $T$ is a translation.

A consequence of this theorem is that the causal structure of Minkowski space or the ensuing relation of temporal precedence $x = y$ completely determines this spacetime including its linear structure and metric, within a scale factor, and is completely independent of the coordinate system or choice of origin [S/Z95, p. 310].

Another consequence is that any causal imbedding of a Lorentzian manifold into another, in particular, any causal automorphism of such a spacetime is in fact a conformal application.

## 5.2 Minkowski spacetime as H(2)

The role of Minkowski spacetime will often be played by the causally isometric real linear space *H(2)*, whose elements are the complex Hermitian matrices of order 2 as we now explain.

We define the real linear isomorphism *J* from $M_0$ to *H(2)* by

$$J(x) = \sum_{k=0}^{k=3} x_k \sigma_k \tag{1}$$

where $x = (x_0, x_1, x_2, x_3)$ and where $\sigma_0$ is the identity matrix and the other three *sigmas* are the Pauli matrices:

$$\sigma_1 = \begin{pmatrix} 0 & 1 \\ 1 & 0 \end{pmatrix}; \ \sigma_2 = \begin{pmatrix} 0 & -i \\ i & 0 \end{pmatrix}; \ \sigma_3 = \begin{pmatrix} 1 & 0 \\ 0 & -1 \end{pmatrix} \tag{2}$$

The set $\{\sigma_k | \ k = 0,1,2,3\}$ forms a linear basis of *H(2)* over the reals. Note that *H(2)* is closed under multiplication by real numbers only: if *H* is Hermitian, *iH* is antihermitian (also called skew Hermitian) [CB/DM/DB77; p.172-173]; [Lev 95, p. 79], [S/Z95, p. 311].

The three Pauli matrices are traceless, Hermitian and of determinant -1. They form a basis for traceless Hermitian matrices..

Hence we have more explicitly [Seg76.1, p. 24, 60]

$$J(x) = \begin{pmatrix} x_0 + x_3 & x_1 + ix_2 \\ x_1 - ix_2 & x_0 - x_3 \end{pmatrix} \tag{3}$$

The closed future cone at the origin of $M_0$ has been defined as

$$\{x = (x_0, x_1, x_2, x_3) \mid x_0^2 - x_1^2 - x_2^2 - x_3^2 \geq 0; x_0 \geq 0\} \tag{4}$$

Under this real linear isomorphism *J* from $M_0$ to *H(2), t*he closed cone field becomes that of positive semidefinite matrices in *H(2,* i.e. matrices *H* such that $X^*HX \geq 0$ for all column complex vectors *X*. The notation $X^*$ denotes the conjugate transpose of *X*. This can be seen as follows. First note that the determinant of the matrix $J(x)$ is precisely $x_0^2 - x_1^2 - x_2^2 - x_3^2$ and that its trace is $2x_0$. Next, thinking of the real diagonal matrix *A* to which $J(x)$ is unitarily equivalent, one sees that the necessary and sufficient condition to have $J(x) \geq 0$, i.e. that $J(x)$ be positive semidefinite, is that the eigenvalues of $A$ both be $\geq 0$ which means that the determinant and the trace both be $\geq 0$. Note that these properties are invariant under conjugations by unitary matrices.

# 6 Causality in the unitary group U(2): the Lie algebra of U(2) as H(2)

The unitary group of 2 x 2 unitary complex matrices is a real Lie group of which the Lie algebra $u(2)$, which is its tangent space at its unit element (the identity matrix) can be identified with $iH(2)$, the real linear space of 2 x 2 antihermitian matrices.

Generally, the group $U(n)$ of unitary matrices of order n has dimension $n^2$; hence $U(2)$ has dimension 4.

To find out which matrices $A$ form the Lie algebra of $U(n)$, one looks for matrices $A$ which generates by exponentiation one-parameter subgroups $\{e^{At} \mid t \in R\}$ of $U(n)$ which are thought of as curves lying in $U(n)$. We obtain successively $e^{At}$ is unitary iff its conjugate transpose is equal to its inverse iff $(e^{At})^* = (e^{At})^{-1}$ iff $e^{A^*t} = e^{-At}$ iff $A^* = -A$ iff $A$ is antihermitian. Such an antihermitian matrix can be written uniquely as $A=iH$ where $H$ is Hermitian [CB/DM/DB77, p. 165, 166], [Mac63, p.32]; [D/N/F82, vol. 1, p. 212].

In the case n = 1 we obtain that the Lie algebra of $U(1)$, which is the unit circle in the complex plane, is the imaginary axis in the same plane.

Usually one takes the real linear space $iH(2)$ of antihermitian matrices as Lie algebra of $U(2)$ as just explained but it is convenient to use the isomorphic real linear space $H(2)$ instead here.

Each Hermitian matrix $H$ defines a tangent vector $X$ at the unit element of $U(2)$ i.e. a linear real valued map on the algebra of analytic real valued differentiable functions $f$ defined in a neighbourhood of the identity element of $U(2)$ and which satisfies the usual differential rule for the product [CB/DM/DB77, p. 117-118]. First, $H$ defines a one-parameter subgroup of $U(2)$ via the exponential map $t$ a $e^{itH}$ which we think of as a curve traced on the 4-dimensional manifold $U(2)$. The real number $Xf$ is the rate of change of the value of $f$ along this curve at $t = 0$ so that $Xf = (d/dt)f(e^{itH})\big|_{t=0}$ [Seg76.1, p, 23; Example 2], [P/S82.1, p. 82].

If $x$ is a point of the manifold $U(2)$, the curve originating at $I$ above can be translated by right multiplication by $x$ to obtain the curve $t$ a $e^{itH}x$ on the manifold $U(2)$ originating at $x$.

The matrices obtained by multiplying the Pauli matrices by $i$ are both antihermitian and special unitary i.e. of determinant 1.

As already mentioned, a basis of the real vector space *H(2)* is made of the three Pauli matrices $\sigma_x = \sigma_1$, $\sigma_y = \sigma_2$, and $\sigma_z = \sigma_3$ together with the identity matrix also denoted $\sigma_0$.

Taking *H(2)* as the tangent space at the identity of *U(2)*, the closed causal cone at the unity element in this tangent space is defined as the set of positive semidefinite Hermitian matrices.

The tangent space and closed causal cone at any other point of *U(2)* is obtained by translation in *U(2)*. We obtain the same cone field no matter whether we use left or right translations in view of the invariance of the future cone in *H(2)* under conjugations by unitary matrices. Hence the cone field on *U(2)* is biinvariant [Seg76.1, p. 61], [Lev 95, p. 80], [CB/DM/DB77] p. 152, 153 ; [N/S79, p. 364-367].

Hence *U(2)* satisfies axioms 1 and 2. It does not satisfy axiom 3, the *globality* condition of the causal structure i.e. there are timelike loops such as $\{e^{it}U \mid 0 \le t \le 2\pi\}$ where *U* is any element of *U(2)*.

# 7   The Hypersphere

## *7.1   Fundamentals on the three-sphere*

Much before Einstein did it in 1917, already in 1854 the very famous German mathematician Bernhard Riemann, in his "*Habilitationschrift*", or inaugural lecture, had envisaged that the three-sphere, also named *hypersphere*, be a model of the universe [Rie1854].

The easiest definition of the hypersphere of radius *r* is given by generalizing the equation of an ordinary two-sphere of radius *r* in three-dimensional space to four-dimensional space. This yields the equation: $x_1^2 + x_2^2 + x_3^2 + x_4^2 = r^2$. In so doing, no physical reality is thereby asserted on the *fourth* space dimension. (It is not denied either.)

To gain an insight into the geometry of this hard to imagine hypersphere it is useful to look at the more familiar 2-sphere. The *parallels* of the surface of the earth idealized as a 2-sphere are circles or 1-spheres; there is one for each latitude. Mentally going from the North Pole towards the South Pole along a meridian, a great circle, one meets these parallels which are increasing in circumference till one reaches the equator and thereafter decreasing to 0 at the South Pole. Similarly, on the three-sphere, going from the North Pole i.e. the point *(0,0,0,r)* towards the antipodal point *(0,0,0,-r)* along any great circle one crosses parallel two-spheres of the three-sphere: there is one for each value of $x_4$ from –*r* to +*r*:   $x_1^2 + x_2^2 + x_3^2 = r^2 - x_4^2$   is the equation of the parallel two-sphere of radius $\sqrt{r^2 - x_4^2}$ . This radius is maximal for $x_4 = 0$ which yields the equatorial two-sphere of radius *r*.

The earth's parallels are ordinary Euclidean circles centered on the axis joining the North Pole to the South Pole. But a parallel may also be viewed as a non-Euclidean circle centered at the North Pole with a *geodesic radius* measured on the earth's surface along a great circle. For instance the equator has *geodesic radius* $\pi r/2$ where $r$ is the earth's radius while the South Pole is a circle reduced to a point but with *geodesic radius* $\pi r$.

Similarly, if our physical space is a hypersphere and one thinks of our position in it as its North Pole and mentally draws concentric geodesic 2-spheres around that position, their surface area would at first increase with increasing geodesic radius till they reach a maximum at the *equatorial 2-sphere* and then contract to 0 at the antipodal point of our position [Rin77, p. 109].

A 2-sphere may be visualized by gluing the edges of two superimposed disks (of equal radii) together and then pushing the upper disk upwards and the lower disk downwards. Similarly, a three-sphere may be imagined as resulting from two ordinary solid balls (superimposed or not, depending on one's preference) whose surfaces are abstractly identified and become the equatorial 2-sphere [Wee02, pp. 201-203].

Astonishingly, Riemann's nineteenth century description of the universe as a three-dimensional sphere appears to have been anticipated much earlier by the Italian poet Dante for whom the universe encompasses the material world as well as Paradise, Inferno and Purgatorio. In his celebrated work '*The Divine Comedy*' (Canto 28, lines 1-129) the thirteenth-century Florentine writer views the Universe from a point in the Primum Mobile [the equatorial 2-sphere] where he stands with his beloved Beatrice who shows him, on the one hand, Paradise which he calls the Empyrean, [an hemi-hypersphere, indeed a three-ball] consisting of a sequence of two-spheres of decreasing radii, lodging angels of all orders, all the way to God [standing at a pole of the 3-sphere] and, on the other hand, the material world [the other hemi-hypersphere; indeed the other three-ball] made of another sequence of two-spheres also of decreasing radii, dwellings of the stars, the planets and the earth with Satan at its centre [the antipodal point of the first pole on the three-sphere], [Pet79], [Oss95, p.89].

More about the three-sphere can be learnt from [Thu97, pp. 32-33 and pp. 103-108].

## 7.2   The three-sphere as SU(2)

Unitary matrices of determinant 1 (unimodular unitary matrices) of order 2 make up the special unitary group *SU(2)*. Generally, *SU(n)* has dimension $n^2 - 1$; hence *SU(2)* has dimension 3.

The unit three-sphere $S^3$ can be identified with the special unitary group *SU(2)* by virtue of the fact that the generic element of that group is the matrix

$$U = \begin{pmatrix} a+ib & c+id \\ -c+id & a-ib \end{pmatrix} \tag{1}$$

where, the determinant being 1, we must have

$$a^2 + b^2 + c^2 + d^2 = 1 \tag{2}$$

which is the equation of the unit sphere in a Euclidean four-dimensional space.

The above matrix $U$ can be written thus in terms of the Pauli matrices together with the identity matrix $\sigma_0$ [S/Z95, p. 311], [Seg93.1, p. 1115], [CB/DM/DB77, p.172-173]:

$$U = a\sigma_0 + i(d\sigma_1 + c\sigma_2 + b\sigma_3) \tag{3}$$

The Lie algebra of the Lie group $SU(2)$ is the set of traceless antihermitian matrices i.e the set of all $iA$ where $A$ is traceless and Hermitian. A linear basis for it is made of the three antihermitian matrices $i\sigma_1$, $i\sigma_2$ and $i\sigma_3$ [CB/DM/DB77, p.173].

This correspondence between $S^3$ and $SU(2)$ endows the former with a group structure and the latter with the Riemannian metric induced on the sphere by the four dimensional Euclidean space of which it is the unit sphere. This metric on $SU(2)$ is the unique one (to within a positive constant multiple) which is biinvariant i.e. invariant under both right and left translations [Seg76.1, p. 61]. Aside from $S^1$, the unit circle, the three-sphere is the only sphere with a Lie group structure.

Left or right multiplication by a fixed element of $SU(2)$ rotates the whole 3-sphere, and any rotation can be produced by a combination of left and right multiplications.

# 8   The relativistic stereographic projection

## 8.1   Remembering the one-dimensional stereographic projection

The relativistic stereographic projection is a generalization of the ordinary concept which we illustrate in one dimension. It is convenient to work in the complex plane. It is algebraically useful, though not geometrically typical, to choose a situation where a circle is stereographically projected on a line which is not tangent to it.

Note though that the chosen straight line is an additive group, and the chosen circle, a multiplicative one. Moreover the straight line, the imaginary axis in the complex plane, is the Lie algebra of the circle which is the unit circle in that plane as we have seen.

The unit circle, minus one point $w' = -1$ is stereographically projected from that point onto the imaginary axis. This maps a point $w$ other than $w'$ on the circle onto the point $z$ on the axis which is the intersection of this axis with the line joining *w* and *w'*. This intersection is defined thus

$$z = (w-1)/(w+1)^{-1} \qquad (1)$$

The inverse mapping maps *z* onto *w* through the following function:

$$w = (1+z)(1-z)^{-1} \qquad (2)$$

These two equations are easily interdeducible and it is best to take the second equation as a definition and obtain the first as a consequence.

As z moves upward all along the imaginary axis, *w* goes around the circle once counterclockwise starting from the point *w'*.

These assertions can be verified using the equation of the circle $w\overline{w} = 1$ and a little complex number algebra: the straight line *D* joining *w'* and *z* which intersects the circle at *w* is obtained by translating the difference vector *z-w'* by *w'*. Hence $D = \{c(z-w') \,|\, c \in R\} + w'$.

### 8.2 The conformal compactification U(2) of Minkowski spacetime H(2)

We will define a causal immersion of *H(2)* into a dense subset of the compact *U(2)* which then acquires the name of *causal* or *conformal compactification* of *H(2)* or of the causally isomorphic $M_0$.

This immersion is the Cayley transform which goes from *H(2)* to *U(2)* and which generalizes the inverse of the stereograhic projection just seen.

This is not the exponential map which also maps the Lie algebra of a Lie group into that group. For instance, the imaginary axis is the Lie algebra of the unit circle *U(1)* and the exponential map covers the circle infinitely many times (this is indeed the projection map from the universal cover of the circle)

One often compares the complex plane with the algebra of operators on a Euclidean or unitary linear space. In this correspondence, in the case of a two-dimensional unitary space, the role of real numbers as elements of the complex plane is played by Hermitian matrices and the role of purely imaginary numbers i.e. matrices *A* such that $A^* = -A$. Just as any complex number is uniquely the sum of a real number and a purely imaginary one, so is any operator uniquely the sum of a Hermitian operator and a skew Hermitian one.

In this generalization of the stereographic projection the role of the imaginary axis will be played by *iH(2),* the algebra of 2 x 2 skew Hermitian matrices and the role of the unit circle will be played by the group $U(2)$ of 2 x 2 unitary matrices. This means that in place of the purely imaginary number $z = bi$ where $b$ is a real number we have a skew Hermitian matrix $iA/2$ where $A$ is Hermitian (the factor 2 is for convenience only); and in place of the complex number $w$ on the unit circle, we have a unitary matrix $U$.

Let us first note that if $H$ is a Hermitian linear operator on a complex inner product space then $I-iH$ is invertible. For otherwise there would exist a nonzero vector $u$ annihilated by $I-iH$ and this would imply, after multiplying $(I-iH)u = 0$ by $i$ that the Hermitian operator would have a non real number, namely $-i$, as eigenvalue [H/K, p.309].
.
Then, starting first with the inverse stereographic projection called the Cayley transform and guided by formula (2) we define:

$$U(A) = (I + iA/2)(I - iA/2)^{-1} \qquad (3)$$

The factor ½ is to ensure that the infinite interval of Minkowskian time going from minus infinity to plus infinity corresponds to an interval of length $2\pi$ which is the time it takes to go around a unit circle at speed 1.

The fact that $U$, i.e. $U(A)$, is unitary i.e. $UU^* = I$ can be seen by noting first that the two matrices $I+iH$ and $I-iH$ where $H = A/2$ commute (the two products yield $I + H^2$) and then computing $UU^*$ remembering that $H = H^*$.

The mapping $A \mapsto U$ admits as inverse mapping i.e. the generalized stereographic projection

$$A = -2i(U - I)(U + I)^{-1} \qquad (4)$$

this being defined for all unitary $U$ such that $U+I$ is invertible. This is obtained by multiplying on the right, both sides of $U = (I + iH)(I - iH)^{-1}$ by $I-iH$ and resolving for $H$ assuming that $U+I$ is invertible. This shows, in particular, that the Cayley transform is one to one.

The set of all unitary matrices $U$ such that $det(U+I) = 0$, a three dimensional submanifold of $U(2)$, replaces the single point $w'$ in the one-dimensional ordinary case. It is the boundary of the image of $H(2)$ in $U(2)$ and is known as the *light cone at infinity* $\mathcal{I}$ (pronounced 'scri' - a contraction of 'script I') [P/R86, p 291].

It is of the utmost importance that the Cayley transform be a causal mapping. In some writings it is the negative of the definition we have chosen that is used equivalently (one being causal iff the other one is). This is so, for instance, on p. 36 of [Seg76.1] where one finds the following theorem of which we present a revised proof. This proof is somewhat different from, and hopefully more comprehensible than, the one presented there and is due to Alexander Levichev.

**Theorem**: The Lie algebra of *U(2)*, as identified with *H(2)* is causally isomorphic to the open dense subset of all unitary matrices in *U(2)* such that *U-I* is invertible via the Cayley transform $U(H) = (H - iI)(H + iI)^{-1}$

**Proof** For the proof, note that the statement means that the future cone at any point *H* of *H(2)* is mapped onto the future cone at the image *U(H)* of *H* in *U(2)*. More explicitly, this means that the image of any tangent vector *F* at *H* is timelike or lightlike iff its image which is a tangent vector at *U(H)* is timelike or lightlike. Recall that the cone at *H* is obtained by translation from the cone at the zero matrix in *H(2)*. Also the tangent cone at *U(H)* is obtained by translation i.e. say, by right multiplication, from the cone at the unit matrix in *U(2)*. Identifying the tangent space at the origin of a linear space with that same linear space we have that a tangent vector at *H* in *H(2)* has the form *H + F* where *F* is also Hermitian.

For $\varepsilon$ larger than or equal to 0, set $H(\varepsilon) = H + \varepsilon F$, which is a tangent vector at the point *H* of *H(2)*. As ε increases from 0, $H(\varepsilon)$ traces a half line originating at *H* in the linear space *H(2)*. Then $U(H(\varepsilon))$ as ε increases from *0* describes a curve in *U(2)* originating at *U(H)*. Right translating this curve by $U(H)^{-1}$ gives a curve $U(H(\varepsilon))U(H)^{-1}$ originating at *I*. The problem is to show that the tangent to this curve at *I*, i.e. for ε = 0, is timelike iff *F* is. To achieve this, one differentiates $U(H(\varepsilon))U(H)^{-1}$ with respect to $\varepsilon$ and evaluates the result for $\varepsilon = 0$. This should give an antihermitian matrix from which the corresponding Hermitian matrix is obtained by dividing by *i*. A calculation shows that this Hermitian matrix is $2(H + iI)^{-1}F(H - iI)^{-1}$.

This expression has the form *G\*FG* where *G* is non singular, and so is Hermitian positive semidefinite iff *F* is.

The computations leading to the antihermitian matrix are as follows. First note the following rule of matrix differentiation where *A* is any invertible square matrix and *A'* its derivative:
$$(A^{-1})' = -A^{-1}A'A^{-1}$$
This rule follows by differentiating both sides of $AA^{-1} = I$ using the ordinary rule for the product.
Then note also that
$$U(H(\varepsilon)) = (H + \varepsilon F - iI)(H + \varepsilon F + iI)^{-1}$$
Now using the fact that $U(H)^{-1}$ is a "constant" independent of $\varepsilon$, one obtains by differentiating $U(H(\varepsilon))U(H)^{-1}$ with respect to $\varepsilon$, and evaluating the result at $\varepsilon = 0$:
$$[F(H + iI)^{-1} - (H - iI)(H + iI)^{-1}F(H + iI)^{-1}](H + iI)(H - iI)^{-1}$$
$$= [F - (H - iI)(H + iI)^{-1}F](H - iI)^{-1}$$
$$= [I - (H - iI)(H + iI)^{-1}]F(H - iI)^{-1}$$
$$= [(H + iI)(H + iI)^{-1} - (H - iI)(H + iI)^{-1}]F(H - iI)^{-1}$$
$$= 2i(H + iI)^{-1}F(H - iI)^{-1}$$

<div align="right">QED</div>

The following theorem is also of the utmost importance

> **Theorem**: Any causal automorphism of $M_0$ extends uniquely to one of $U(2)$ in such a way that the group of causal automorphisms of $M_0$ becomes a subgroup of that of $U(2)$. [P/S82.1, lemma 2.1.3, p. 85], [Seg93.1, Theorem1, p. 11114]

There is another approach to the compactification of Minkowski spacetime. Instead of generalizing the stereographic projection it generalizes another way of compactifying which is that employed in the definition of the projective line or, more generally, of projective n-space $P^n$ which compactifies Euclidean n-space $E^n$ by identifying antipodal points in the n-sphere $S^n$. In this approach, compactified Minkowski spacetime appears as a projective quadric in 6-dimensional space. [P/R86, p. 297-303]. We will not develop this here. From this approach it is easy to see that the conformal group is $O(4, 2)$ which is of dimension $15 = (2 + 4)(2+4-1)/2$ [Seg76.1, p. 6].

The generalized stereographic projection has an effect similar to the ordinary geographical one used in world maps : it makes far away regions look much larger than they are in fact. If we look far in the future or in the past i.e. if we let cosmic time $s$ approach $\pi$ or $-\pi$, (or $\pi r$ and $-\pi r$, if $r$ differs from 1) spacetime regions tend to look infinitely larger in all directions of spacetime when expressed in Minkowskian coordinates. Spacewise this is the neighborhood of our antipode viewed in the Minkowskian nearly infinite future or in the Minkowskian nearly infinite past. This is similar to the familiar situation which makes the neighborhood of the North Pole look larger than reality on plane maps of the northern hemisphere of the earth.

### 8.3 *The light cone at infinity*

In this section we adopt the traditional definition-theorem style.
Let us recall the formulae defining the immersion of $H(2)$ into $U(2)$ by the Cayley transform and its inverse defined on the image of the transform:

$$U(A) = (I + iA/2)(I - iA/2)^{-1} \qquad (3)$$

$$A = -2i[U(A) - I][U(A) + I]^{-1} \qquad (4)$$

We have that $det(U(A)+I) \neq 0$ for all $A$ in $H(2)$ since the Cayley transform is invertible on its image.

> **Theorem 1**: $det(U(A)-I) = 0$ iff $detA = 0$ i.e. $A$ is a light ray at the origin of $M_0$.

> **Definition 1**: The light cone at the origin in $U(2)$, $LCOU(2)$) is the set of all $U$ in $U(2)$ such that $det (I-U) = 0$.

One may prove that:

$$LCOU(2) = \{U \in U(2) \,|\, \det(1-U) = 0\} = \{U \in U(2) \,|\, \det(U) - tr(U) + 1 = 0\} \qquad (5)$$

This includes by Theorem 1, the image in $U(2)$ of the light cone at the origin of $M_0$ via its representation in $H(2)$.

> **Definition 2** The *light cone at infinity in U(2), LCIU(2)* is the set of all $U$ in $U(2)$ such that $det(I+U) = 0$. It is the complement, indeed the boundary, of the image of $H(2)$ in $U(2)$ by the Cayley transform.

One may show that

$$LCIU(2) = \{U \in U(2) \,|\, \det(1+U) = 0\} = \{U \in U(2) \,|\, \det(U) + tr(U) + 1 = 0\} \qquad (6)$$

These descriptions of *LCIU(2)* and *LCOU(2)* can be established by using the characteristic polynomials of $I+U$ and $I-U$.

As we have noted, every causal automorphism of $M_0$ extends uniquely to $U(2)$. However there are causal automorphisms of $U(2)$ under which $M_0$ is not closed. Also there are conformal transformations not everywhere defined in $M_0$ which extend uniquely to everywhere defined conformal transformations on the larger space $U(2)$.

Here is a first example. In Minkowski spacetime *conformal inversion* is defined [Seg76.1, p. 71] thus $Q : X \to 4X / X^2$ (the factor 4 is for convenience only). This is a mapping which is undefined on the light cone at the origin where $X^2 = 0$. In $H(2)$ this is the set of matrices of determinant $0$. [D/N/F82, vol.1, pp. 139-146]

This partial mapping extends to a genuine transformation of $U(2)$ still denoted by $Q$ once $H(2)$ is identified with a submanifold of $U(2)$ via the Cayley transform [Seg76.1, p. 71, 72].

> **Theorem 2:** The action of conformal inversion $Q$ in $U(2)$ is given by $Q(U) = -U / \det(U)$ [Seg76.1, p.72]

Thus $Q$ is really a causal automorphism of $U(2)$ under which $M_0$ (as a subset of $U(2)$ by way of the Cayley transform) is not closed.

> **Theorem 3** $Q^2$ acts as the identity mapping on $U(2)$.

> **Theorem 4:** The image of *LCOU(2)* by conformal inversion is *LCIU(2)* and hence conversely by virtue of Theorem 3. In other words, conformal inversion maps the light cone at the origin and the light cone at infinity on each other in $U(2)$.

Thus we have

**Corollary**: The *light cone at infinity in U(2), LCIU(2)* includes the image by conformal inversion *Q* of the image of the light cone at the origin of *H(2)* by the Cayley transform.

One can see easily from their above descriptions that the intersection of *LCIU(2)* and *LCOU(2)* is the following set

$$\{U \in U(2) \,|\, \det(U) = -1 \text{ and } tr(U) = 0\} \tag{7}$$

The condition *det(U) = -1* yields immediately the fact that the intersection is pointwise invariant under conformal inversion. This shows in particular that

**Theorem 5**: *LCOU(2)* includes matrices which are not in the light cone at the origin in *H(2)*.

Alex Levichev, has shown us the following

**Theorem 6**: The intersection of *LCIU(2)* and *LCOU(2)* is the following 2-sphere made of the elements *U* of *U(2)* which can be written

$$U = \begin{pmatrix} -z & -y - ix \\ -y + ix & z \end{pmatrix} \tag{8}$$

in which *x*, *y* and *z* are real numbers and where

$$x^2 + y^2 + z^2 = 1. \tag{9}$$

**Proof**. Indeed, the condition that *tr(U) = 0* and the fact that *U* is unitary of order 2 imply that the eigenvalues must be real and the negatives of each other. This implies in turn that the matrix must be Hermitian and hence (8). The condition that *det(U) = -1* yields (9).

<div align="right">QED</div>

## 8.4   The two groups of temporal translations in U(2)

Another important example of causal automorphisms of *U(2)* under which $M_0$ is not closed is the one-parameter group which Segal calls the *unitime group* [Seg76.1, p. 70]. For a given real number *t*, this maps *U* onto *exp(it)U*. This parameter group will turn out to be the projection on *U(2)* of the cosmic time parameter group in the Einstein universe.

For instance letting $t = \pi$ and *A* in *H(2)* be of determinant *0*, i.e. *A* is a null vector as an element of $M_0$, we have that $e^{it}U(A) \notin M_0$, i.e. not in the image of $M_0$ in U(2) since then $e^{it}U(A) = -U(A)$ and $\det[I + (-U(A))] = 0$ by virtue of Theorem 1 of the last section.

Therefore, for this $t$ and this $A$, $e^{it}U(A)$ belongs to the light cone at infinity by definition of *LCIU(2)* which is the complement of the image of $M_0$ in U(2)  The unitime group is therefore distinct from the temporal translation group in $M_0$ which extends to a causal group of automorphisms of *U(2)*.

So the unitime group maps $U$ in $U(2)$ onto *exp(it)U* and, on the other hand, the Minkowskian time translation group maps $U(A)$ onto $U(A + tI)$ for $A$ in $H(2)$.

These two non conjugate subgroups  of the conformal group acting on U(2), differ very little in the neighborhood of the identity matrix I representing in $U(2)$ the origin of Minkowski space.  Indeed, for a given value $t$ of the parameter, these two groups map  $I$ respectively as follows:

- $I \rightarrow e^{it}I$, for the unitime group

- $I \rightarrow (1+it/2)(1-it/2)^{-1}I$  for the Minkowskian time translation group

These differ by $O(t^3)$. [Seg76.1, p. 70]

# 9   The standard causal imbedding of Minkowski spacetime in the Einstein universe

## 9.1   The Einstein universe as the universal covering group of U(2)

Let us first review some fundamental concepts of point set topology.  A continuous surjective map  $p : Y \rightarrow X$  between two topological spaces is said to be a *covering projection* if any point in $X$ has an open  neighborhood $U$ such that the inverse image of $U$, $p^{-1}(U)$ is the disjoint union of open subsets of $Y$ such that each of them is mapped homeomorphically onto $U$ by $p$.  Then $Y$ is called a *covering space* of the *base space X*. Any one of these open subsets of $Y$ is called a *section* over $U$ and the inverse homeomorphism from $U$ to that subset is called a *section map* over $U$ [Spa66, p. 62].

It follows that a covering projection is a *local homeomorphism*, i.e. each point of its domain $Y$ has an open neighborhood which is mapped homeomorphicaly onto its image [Spa66, p. 63].

The number of elements of $Y$ mapped on a given element of $X$ is independent of that element and is called the *multiplicity* of the covering.  This number may be finite or infinite.  If this number is $c$ we say that the covering is *c-sheeted* [Wolf, p. 35].

In any projection covering  $p : Y \rightarrow X$, if the base space $X$ is endowed with a differential structure, even a causal structure, such structure can be *lifted* to the domain $Y$ of $p$ uniquely so that the map $p$ preserves the structure [Var84, p.65], [Wolf84, p. 41], [N/S79, p. 370, VI].

If the base space is endowed with a group structure, this can be lifted to the covering space, but this time only to within isomorphism. Any element of the covering space mapped onto the identity element of the base space by the covering map may be selected as the identity element of the cover. The covering map then becomes a group epimorphism [N/S79, p. 333], [Che46, p. 53].

A first pertinent example is the exponential map $ex : R \rightarrow S^1$ from the real line to the unit circle in the complex plane defined as $ex(t) = e^{2\pi i t}$.

A covering projection $p : Y \rightarrow X$ is said to be universal if Y is a simply connected topological space which is then called a universal covering space.

Under some conditions a topological space admits a unique topological covering space.

The unit circle is not simply connected but the real line $R$ is (N/S79, p. 335). One says then that $R$ is the universal covering space of $S^I$. The covering projection being the above exponential map.

Here is another example of covering projection. It maps the group $S^I \times SU(2)$ onto $U(2)$ thus: $(e^{i\theta}, V)$ a $e^{i\theta}V$. This is a double (or two-sheeted) covering of $U(2)$ : each element of $U(2)$ has two inverse images since $(e^{i\theta}, V)$ and $(e^{i(\theta+\pi)}, -V)$ have the same image.

Note that $SU(2)$ is closed under the taking of the additive inverse since 2 is even and hence the determinant of an element of $U(2)$ is the same as that of its additive inverse.

This two-sheet covering is not universal because $S^I$ is not simply connected.

The sphere $S^n$ is simply connected for all $n$ larger than 1. Hence $SU(2) = S^3$ is simply connected [Spa66, p. 58]. Also $SU(n)$ is simply connected for all n. [Che46, Prop. 6, p. 60].

The product of simply connected spaces is simply connected [N/S79, p. 326], [Che46, p. 45, prop. 1].

It follows that the universal covering of $U(2)$ is $R \times S^3$ or, equivalently, $R \times SU(2)$.

The covering map is $(t, V)$ a $e^{it}V$ where $t$ is any real number and $V$ any matrix in $SU(2)$.

Note that $(t + 2\pi, V)$ and $(t + \pi, -V)$ have the same image as $(t, V)$.

It follows that the covering projection $R \times SU(2) \rightarrow U(2)$ has an infinite number of sheets.

As already said, the Lorentzian metric on the Einstein universe $M = R \times S^3$ is the hyperbolic product $c^2 dt^2 - r^2 ds^2$ of the ordinary metric on $R$ with the ordinary metric on $S^3$. This metric on the sphere corresponds to a unique biinvariant metric on the group $SU(2)$. The metric on $M$ is biinvariant . It defines the same cone field on $M$ as that obtained by lifting the one from $U(2)$. [Lev95, p. 81], [S/Z95, p. 310], [Seg76.1, p. 61]

The isometry group $K$ on the Einstein universe can be described in two equivalent ways; first as the set of all products of a time translation with a space rotation [S/Z95, p. 311]; second as the group of all right or left translations of the group structure of the Einstein universe. The rotations of the 3-sphere are the same as products of right or left translations on $SU(2)$. It is a seven parameter subgroup of the fifteen parameter causal group of $M$. As already said, there is in general no isometry which will transform one given timelike direction into another [Seg76.1, p. 59]. The same can be said of spacelike directions [Tits60, p. 112].

## 9.2   *Lifting $M_0$ to a section of its universal cover*

Any element  $U$ of $U(2)$ may be expressed in the form $exp(it)V$ where $t$ is a real number in the semi-open interval $(-\pi, \pi]$ and $V$ belongs to $SU(2)$ and is unique. The parameter $t$ in this expression is made unique by the constraint that the map which assigns $t$ to $U$ be continuous and that  $t = 0$ if $U = I$.  [S/Z95, p. 312 top].  [For a fixed $V$, the determinant of  $U = exp(it)V$ goes around the unit circle twice as $t$ runs through the interval  $(-\pi, \pi]$ once.]

The mapping $U$ a  $(t,V)$  from $U(2)$ to $R \times SU(2$ yields a section[2] of the covering.  If this is preceded by the causal immersion of $H(2)$ into $U(2)$ we obtain a causal immersion of $H(2)$ into $R \times SU(2)$. If we replace $H(2)$ by $M_0$ and $SU(2)$ by $S^3$ we thus obtain a causal immersion of Minkowski spacetime into Einstein spacetime through the following sequence of mappings

$$M_0 \rightarrow H(2) \rightarrow U(2) \rightarrow R^1 \times SU(2) \rightarrow R^1 \times S^3 = M \qquad (*)$$

By virtue of properties of universal covering spaces, causal automorphisms of $U(2)$ can be uniquely lifted to causal automorphisms of the universal covering space $M$. [P/Z82, lemma 2.1.3, p. 85], [Seg93.1, Theorem1, p. 11114] , [S/Z95, p. 318]

For instance, the unitime translation group in $U(2)$ lifts to the cosmic temporal translation group in $M$.  The conformal inversion in $U(2)$, $Q(U) = -U / \det(U)$ lifts to $Q^\infty (t,V) = (\pi - t, V)$ in the infinitely-sheeted $R \times SU(2)$.

---

[2] This immersion is continuous except at points where $t = \pi$; it is continuous at all points of $M_0$.

# 10 The Minkowskian-Cosmic times formula of Chronometric cosmology

We are now in a position to present here a proof of the *Minkowskian-cosmic times formula* and the ensuing *redshift formula of CC* mostly based on [Seg93.1]. These formulas, already mentioned in our introduction, are the following:

$$x_0 = 2r \tan(s/2r) \tag{1}$$

$$z = \tan^2(s/2r) \tag{2}$$

in which $x_0$ is Minkowskian or local time, $s$ the Einsteinian or cosmic time, $r$ the radius of the 3-sphere and $z$ is the observed redshift after a time of propagation $s$ along a geodesic of the 3-sphere.

For simplicity's sake, we assume at first, the three-sphere radius $r$ to be unity. As we also assume the speed of light $c$ to be 1, the geodesic distance traveled on the three-sphere is equal to the cosmic time elapsed. The formula (1) to be established then becomes simply $x_0(s) = 2 \tan(s/2)$ in which $s$ is cosmic time and $x_0(s)$ is the corresponding Minkowskian time. The roles of the two main characters in this drama, namely Minkowski spacetime $M_0$ and the unit three-sphere $S^3$, will be played respectively by $H(2)$, the set of $2 \times 2$ complex Hermitian matrices, and $SU(2)$, the special unitary group made up of the $2 \times 2$ unitary matrices of determinant 1 as we have explained previously.

In what follows we identify $M_0$ with $H(2)$ and $M$ with $R^1 \times SU(2)$ keeping in mind the already detailed four-step imbedding of $M_0$ into $M$ summarized thus

$$M_0 \rightarrow H(2) \rightarrow U(2) \rightarrow R^1 \times SU(2) \rightarrow R^1 \times S^3 = M \tag{3}$$

Hence each point in $M_0$ has four names: one $x$ in $M_0$ proper, an Hermitian matrix

$$A = \begin{pmatrix} x_0 + x_3 & x_1 + ix_2 \\ x_1 - ix_2 & x_0 - x_3 \end{pmatrix} \tag{4}$$

a unitary matrix

$$U = (I + iA/2)(I - iA/2)^{-1} \tag{5}$$

and *(t, V)* as a member of $M$ where $U = e^{it}V$.

The stage is now set for the proof of the two-times formula.

One must think of a photon being emitted somewhere on the three-sphere and being observed later elsewhere after some cosmic time $s$. It is important to distinguish between three events all of which belonging to the image of $M_0$ in $M$ as we suppose that the photon is observed after less than one half-tour of the three-sphere merry-go-round which it is circling along a grand circle. The three events are: the emission *(t, W)* of the photon at cosmic time *t* at the point *W* of space *SU(2)*; the observation *(t + s, V)* of the photon *s* cosmic time later at the point *V* of *SU(2)*; and the event *(t, V)* when a patient observer starts waiting at *V* for the arrival of the photon at the moment it is emitted with plenty of goulash at her side to keep her alive and in good spirits long enough !

Let *A* be the Hermitian matrix corresponding to the event *(t, V)* with $t = x_0$ and *U* be its Cayley transform defined by (5). Let *A(s)* be the Hermitian matrix corresponding to the event *(t + s, V)*. One obtains from the already noted formula

$$A(s) = -2i(e^{is}U - I)(e^{is}U + I)^{-1} \qquad (6)$$

This follows from the fact that the Einsteinian temporal translation $T_s$, the isometry of *M* which maps any *(t, V)* onto *(t + s, V)*, once interpreted in the notation of *U(2),* maps any *U* onto $U(s) = e^{is}U$. From this, one obtains after some calculations

$$A(s) = 2(2aI + bA)(2bI - aA)^{-1} \qquad (7)$$

where $a = \sin(s/2)$ and $b = \cos(s/2)$.

We may safely assume that the observation takes place at the origin of $M_0$, and that the cosmic time of emission is *t = 0* so that *A = 0, U = I = V*. It then follows immediately from (7) that

$$A(s) = 2\tan(s/2)I \qquad (8)$$

By definition the matrix A(s) can be written

$$A(s) = \begin{pmatrix} x_0(s) + x_3(s) & x_1(s) + ix_2(s) \\ x_1(s) - ix_2(s) & x_0(s) - x_3(s) \end{pmatrix} \qquad (9)$$

In view of the fact that $x_j(s) = 0$ for *j = 1, 2, 3,* one obtains from (8) and (9) the desired conclusion $x_0(s) = 2\tan(s/2)$.

A better understanding of the two-times and the redshift formulae is achieved at the cost of intensifying the calculations. This consists in looking at the effect of the temporal translation $T_s$ on some small neighborhood *NGR* (in the image of $M_0$ in *M*) of the origin of $M_0$ instead of this effect just on the origin of $M_0$. This is done through the differential approximation $dT_s$ of $T_s$. For a point $x = (x_0, x_1, x_2, x_3)$ other than the origin in *NGR* we no longer have *A = 0* nor $x_j(s) = 0$ for *j = 1, 2, 3* but we still have (6), (7) and (9).

The linear transformation $dT_s$ maps the tangent space at the origin of $M_0$ (as imbedded in *M*) onto the tangent space at the image of the origin by $T_s$. The matrix of this linear transformation expressed in Minkowskian coordinates is the *Jacobian*

$$\left( \frac{\partial x_i(s)}{\partial x_j} \right) \qquad (10)$$

evaluated at the origin of $M_0$. A computer calculation shows that this is the diagonal matrix all elements of the main diagonal being equal to $\sec^2(s/2)$ [3]. This means that the approximate effect of $T_s$ on any point $x$ of $NGR$ is to map it on a point whose Minkowskian coordinates are those of $x$ magnified by the same factor $\sec^2(s/2)$. Perhaps surprisingly, this is reminiscent of big bang cosmology   except that here Minkowskian time also is expanding. This does not contradict the fact that $T_s$ being an isometry of $M$, it maps $NGR$ isometrically onto its image. Wavelengths which are small relative to the radius of the universe may be assumed to fall within $NGR$. As a result, Segal concludes [Seg93.1, p. 11115]: "In particular, wavelengths, after time $s$ are observed as magnified by the factor $\sec^2(s/2)$."

Thus if $\lambda$ is such a wavelength, it becomes $\sec^2(s/2)\lambda$ under $T_s$. As a result one obtains a proof of the redshift formula using the definition $z = \Delta\lambda/\lambda$ as follows

$$z = \frac{\sec^2(s/2)\lambda - \lambda}{\lambda} = tg^2(s/2) \qquad (11)$$

Similarly a small interval $ds$ of cosmic time is magnified by a factor of $\sec^2(s/2)$ into a larger interval $dx_0(s)$ of Minkowskian time under the temporal translation $T_s$ so that $dx_0(s)/ds = \sec^2(s/2)$. Integrating this relation immediately yields the two-times relation $x_0(s) = 2\tan(s/2)$ taking into account the initial condition $x_0(0) = 0$.

Going back now to the general case of a three-sphere of any radius $r$ instead of the unit sphere, we note that all but the last map in (3) remain unchanged whereas the last one, $R^1 \times SU(2) \to R^1 \times S^3$, must be multiplied by $r$. As a result both $x_0$ and $s$ must be divided by $r$ in the formulae we have just established in the case $r = 1$. This immediately yields the original chronometric two-times and redshift formulae (1) and (2).

# 11 Two significant mathematical facts about the Einstein universe

They may impress mathematicians and particularly mathematical logicians more than astrophysicists and astronomers.

---

[3]  Jean-Marc Terrier, an expert in the computer program *Mathematica*, has kindly verified this result as well as the formula (7). The formulae for $x_j(s)$, $j = 0,1,2,3$, given on p. 11115 of [Seg93.1] appear to be wrong but, in any case, they are not needed here.

### 11.1 It is the universal cosmos

Minkowski spacetime and the Einstein universe are two Friedmann-Lemaître-Robertson-Walker (FLRW) spacetimes which are generally used as cosmological models of the universe. The latter has the distinguishing property that all others can be conformally imbedded into it by essentially unique causality-preserving maps, though the metric of time or space, and the factorization of spacetime into time and space, may not be preserved by such imbeddings. In particular, topologically, space in the imbedded spacetime may not be compact. This property bestows on $M$ the name of *universal cosmos*. To a mathematician's mind, this fact alone gives $M$ a very special status and calls it to his or her attention. But admittedly, this may leave most astrophysicists indifferent [Dai05], [H/E73, p. 139], [Pen67, p.197].

All FLRW spaces appear as open submanifolds of $M$ which are orbits of some subgroups of the conformal group in the Einstein universe. For instance de Sitter spacetime, whose isometry group is $SO(1,4)$, is the orbit in $M$ of any of its points under the action of this subgroup of the conformal group [Seg84.2, p. 146], [Seg90, p. 161].

As we have seen, Minkowski spacetime also illustrates this fact:

> *"For all intents and purposes, $M_0$ as a causal manifold (e.g., as regards light propagation) may be regarded as a submanifold of $M$ that is uniquely determined by the scale extended Poincaré subgroup of the conformal group of which it forms an orbit. $M_0$ inherits the causal structure of $M$ and enjoys just those symmetries that are restrictions of those of $M$."* [S/W95, p. 318]

### 11.2 Maxwell's equations belong to it

Maxwell's equations govern light and more generally electromagnetic radiation, which is basically all that is observable in large scale astronomy. They can be stated in several ways. Their most telling and concise appearance uses the modern language of differential forms.

The differential approach to Maxwell's equations, [MTW73, p. 99-110], is to abstract the electromagnetic field as a 2-form, i.e. an antisymmetric second-rank tensor $F = \frac{1}{2} F_{\alpha\beta} dx^{\alpha} \wedge dx^{\beta}$ where the $F_{\alpha\beta}$ are the electric and magnetic field strengths ($F$ as in Faraday) which are then unified in the field tensor.

Written more explicitly in a somewhat different notation this 2- form looks as follows

$$F = E_x dx \wedge dt + E_y dy \wedge dt + E_z dz \wedge dt + B_x dy \wedge dz + B_y dz \wedge dx + B_z dx \wedge dy \qquad (1)$$

where $(E_x, E_y, E_z)$ and $(B_x, B_y, B_z)$ are the usual electric and magnetic vectors. In this abstract approach, Maxwell's equations take the very simple form

$$dF = 0 = \delta F \tag{2}$$

where $d$ denotes the usual differentiation operator on forms and $\delta$ is its adjoint $d*$ with respect to the adjoint operator *star* on forms, corresponding to the Minkowski metric. [Seg76.2, p. 671], [Seg84, p. 245], [MTW, p. 99, (4.4)], [Thi79, pp. 29, 30]

Genesis teaches that : « …et Dieu dit: ' Que la lumière soit !'  Et la lumière fut. » ; « …and God said : 'Let there be light !'  And light there was ».

Some say that at MIT, Caltech, and other trade schools, one often sees engineers wearing T-shirts that display Maxwell's equations as God's real words just before there was light [ http://wiki.yak.net/591/howto.pdf].

Maybe this is so for the following facts which do not seem to move astrophysicists should not leave any mathematician indifferent especially those like myself who believe that God is Himself a mathematician !

Maxwell's equations $dF = 0 = \delta F$ are invariant under all causal, transformations.  This means that they are the same whether they are computed with respect to the Minkowski metric g or to any other g' yielding the same causal structure i.e. which is conformally equivalent to it:  $g'(x) = \exp(\Phi(x))g(x)$ for some smooth strictly positive real valued function $\Phi(x)$. [Seg84, p. 248-249], [CB/DM/DB77, p. 339-340]

Moreover, every free photon wave function in $M_0$ extends uniquely to a solution of Maxwell's equations throughout $M$, and every solution of Maxwell's equations in $M$ arises in this way. [S/Z95, p. 316], [Seg93.2, p. 4805]

A sketchy proof can be found in  [P/S82, p.  404-408] and [Seg86, p. 219-220].

Also, a somewhat incomplete proof of the redshift formula on the basis of Maxwell's equations is presented in [S/Z95, pp. 318-319] as a rigorous mathematical consequence of these equations.

# 12 Energy conservation and the axiom of causal temporal homogeneity

## 12.1 The principle of energy conservation

As Roger Penrose says, "The utility of the concept of energy, in general, arises from the fact that it is *conserved*." [Pen67, p. 171]

The *integral* (sometimes called local) energy conservation law asserts that the total energy in a given *bounded region of space* varies with time only as a result of the energy gained or lost through the boundary of this region [Fey64, p. 27-3].

This can be rephrased to mean that the total *flow of energy* (a vector field) through the boundary of an *oriented compact spacetime region* vanishes. It suffices to assert this for domains that are bounded regions of space over a finite interval of time [S/W77, p. 62, 63], [H/E73, p. 62].

This does imply, and indeed is equivalent to, a *differential* energy conservation law [S/W77, p. 64 bottom]. The (infinitesimal or) differential energy conservation law, asserts that the divergence of the flow of energy vanishes at each point of spacetime. This means that the rate of loss or gain of energy per unit volume and unit time at this point vanishes. [Swokowski, Calculus, p. 1220] This is generally demonstrated for special relativity i.e. in Minkowski spacetime only and no generalization of this law is known in general relativity [S/W77, p. 62-65, 96, 97], [M/T/W73, p. 142- 146], [H/E73, p. 62 bottom]. This equivalence is demonstrated using Stoke's theorem (also known as Gauss's theorem or as the divergence theorem).

> *"It is a shame to lose the special relativistic total energy conservation law in general relativity."* [S/W77, p. 98]

This is for lack of a concept of flow of energy in GR:

> *"Physically observable energy conservation requires in general the existence of a one-parameter group of temporal isometries , whose generator then corresponds to the conserved energy, and there is no such group in FLC"* (From preprint of [S/W95, p. 11])

Energy conservation is valid in both $M$ and in $M_0$.

In both cases energy is the generator of the temporal translation group which is not only a causal group but a group of isometries.

> *"This defines both the classical and the quantum energy."* [Seg80, p.391]

## 12.2 Hamiltonian systems in classical mechanics

The first mathematical structure where the principle of conservation of energy appears is that of Hamiltonian systems in classical mechanics. The following section is meant to be a short reminder of energy conservation in that context.

In general a *reversible* physical system is specified by a *state space S* and a one-parameter group $U_t$ of transformations on that space called the *evolution group* or *dynamical group* of the system. The interpretation is that $S$ is meant to be the set of all possible states in which the system may be and $U_t(x)$ is supposed to be the state of the system after $t$ units of time have elapsed since it was in state $x$, for any $x$ in $S$. If $t$ is negative, one should rephrase this appropriately. The group $U_t$ partitions the state space into orbits along which the system evolves. Often the state space is endowed with some structure which is preserved by the evolution group [Mac63, p. 1].

A Hamiltonian system is such a reversible physical system specified by a differential manifold $M$ called its configuration space together with a real valued function called the Hamiltonian $H$ (the energy) on the cotangent bundle of this space. This bundle is the state space of the system. The differential form $dH$ on the phase space corresponds canonically to a vector field on the tangent bundle of phase space, called a Hamiltonian vector field which is the one definable by Hamilton's equations of motion [Arn78, p. 203, 207; p. 65].

This is because the cotangent bundle on any differential manifold has a canonical structure called a symplectic structure which allows an abstract formulation of Hamilton's equations. This symplectic structure establishes a canonical isomorphism between tangent vectors and differential forms on phase space or on any manifold equipped with such a structure.

This vector field in turn gives a one-parameter group of diffeomorphisms on phase space called the Hamiltonian phase flow of the system and which is its evolution group [Arn78, p. 204].

At any given point $x$ of phase space the vector field measures the speed at which the system evolves along the orbit, known as a *flow line*, defined by the group and going through that point. The vector field, and by extension the Hamiltonian, is called the (infinitesimal) *generator* of the evolution group.

Energy conservation means that the value of the energy $H$ remains constant along any orbit of the flow [Arn78, p. 207].

The evolution group preserves the symplectic structure of phase space [Arn78, p. 201-204].

In the more classical framework sometimes called *mechanical* or *natural* [Arn78, pp. 66, 84], the manifold $M$ is assumed to be Riemannian. Then the metric establishes an isomorphism between tangent vectors and differential forms on configuration space $M$ and state space can then be taken to be the tangent bundle on $M$ in place of the cotangent bundle. The mechanical or natural Hamiltonian function $H$ defined on the tangent bundle is assumed to be of the form $H = T + U$ where $T$ is called the kinetic energy $U$ the potential energy. The former is the quadratic function defining the Riemannian metric on $M$ and $U$ depends only on the coordinates of $M$.

The Hamiltonian vector fields (or equivalently, the Hamiltonian functions modulo the constant ones) form an infinite dimensional Lie algebra which belongs to the infinite dimensional Lie group of diffeomorphisms preserving the symplectic structure [Arn78, p. 208-218].

### 12.3 Shroedinger's equation in quantum mechanics

The dynamical group of a quantum mechanical system is also generated by a Hamiltonian which takes the form of a Hermitian operator $H$. The group is the one-parameter group of unitary operators $U_t = e^{-itH}$ acting on the complex Hilbert space whose rays represented by unit vectors $\phi$ are the states of the system. As time evolves, the state $\phi$ becomes $\phi_t = U_t(\phi) = e^{-itH}(\phi)$ from which the celebrated Schrödinger equation follows by differentiation with respect to t [Var85, p. 291], [Mac78, p. 169]:

$$d\phi_t / dt = -iH(\phi_t).$$

Limiting ourselves to free relativistic quantum mechanical particles, the system is described by a unitary representation of the (connected component of the identity) of the spacetime relativity group i.e. the Poincaré group. The dynamical group of the system is the restriction of this representation to the temporal evolution subgroup of the Poincaré group [Var85, p. 328-329], [Mac78, p. 168].

The expected value of the energy of the system in the state $\phi$ is given by the expression $< H\phi, \phi >$ which remains constant on any orbit since the evolution group is one of unitary operators which commute with $H$. This is the quantum mechanical form of the "law of conservation of energy" [Mac78, p. 179].

### 12.4 Differential energy-momentum conservation in GR

In general relativity an *energy-momentum* (or *stress-energy*) tensor $T^{ab}$ field is postulated whose divergence (also called its *exterior derivative or gradient*) is assumed to be 0 yielding a differential (or infinitesimal) energy-momentum conservation law [S/W p. 97], [H/E, p. 61, 3.1], [M/T/W73, p 146, 5.36, p. 80-82, 208-212, 259], [Arn78, p.188-200], [S/Wp.71].

By definition, the tensor $T$ is a symmetric real valued bilinear function of two variables each of which is a 1-form which is positive semidefinite [S/W77, p. 71].

But *'no honest integral conservation law is in general implied'* [S/W, p. 87, lower part] . There is in general no natural way to integrate the stress-energy tensor or to interpret the fact that its divergence is 0 via integral conservation laws [S/W, p. 96 middle].

For this, *a Killing vector field* is needed to extract from the energy-momentum tensor a vector field interpreted as the *flow of energy* [H/E, p. 62].

If the group of diffeomorphisms generated by a vector field in a Riemannian manifold is a group of isometries, this vector field is called a Killing vector field [H/E, p. 43, 62].

For any conservation law, a Killing vector field is needed to extract from the energy-momentum tensor, a vector field of zero divergence [H/E, p. 62 upper part], [S/W, p. 96, Prop. 3.10.1] such as the flow of energy, for which the differential and integral conservation laws are valid. There are nine other conservation laws in special relativity [H/E, p. 62].

The gradient raises the rank of a tensor by 1. [Arn78, p.188-200]

## 12.5 Energy conservation in Minkowski and Einstein spacetimes

Minkowski spacetime and the Einstein universe both have the required temporal group of isometries for energy conservation contrary to other GR cosmological models.

In Minkowski spacetime this temporal group is simply given by the mapping $(x_0, x_1, x_2, x_3)$ a $(x_0 + t, x_1, x_2, x_3)$. Upon identifying Minkowski spacetime with $H(2)$ this becomes $H$ a $H + tI$ .

This is a subgroup of the causal group of $M_0$ which is a subgroup of the conformal group after the identification of $M_0$ as a subspace of $M$.

In $U(2)$, the temporal group is $U$ a $e^{it}U$ . This is lifted to the Einstein universe $M = R \times S^3$ into the group $(t_0, p)$ a $(t_0 + t, p)$

Thus both groups are subgroups of the causal group of $M$ and hence both preserve Maxwell's equations which are invariant under all causal transformation local or global [Seg80, p. 391].

The causal group of $U(2)$ is $SU(2,2)$ and the infinitesimal generators of the two groups of temporal translations in $U(2)$ can be identified in the Lie algebra su(2,2). [Seg76.1, p. 70], [S/W95, p. 311]

The flux (or flow) of energy is the vector which is the component of the energy-momentum tensor along the generator $K$ of the temporal evolution group in special relativity or in Einstein universe. It can be defined as the vector $P^a = T^{ab}K_b$ from the energy momentum tensor $T^{ab}$ and the Killing vector $K = d/dx^0$ or $K = d/dt$ generating the temporal translation group. The fact that $T^{ab}$ is symmetric and of zero divergence and the fact that $K$ is Killing imply that the divergence of this flow of energy also vanishes which means differential energy conservation. From this, one obtains the integral law of energy conservation using Stokes theorem as said above [H/E73, p. 62]; [S/W73, p. 96, Prop. 3.10.1].

We get two different concepts of energy, the Minkowskian or local energy and the Einsteinian or cosmic energy, depending on which temporal group we use.

The cosmic energy exceeds the local one; the difference, as far as photons are concerned, is the redshift.

The conventional formula $h\nu$ for the energy of the photon is only the local or Minkowskian energy. It measures the true i.e. cosmic energy only in the case of freshly emitted photons in which case the two energies differ very little.

The energy that is conserved is, in particular, that of electromagnetic radiation in vacuum "which is basically all that is observable in large scale astronomy." Both electromagnetic energies, the Minkowskian and the Einsteinian, are conserved in their respective cosmos $M_0$ and $M$:

> *"They are both representable as the integral over space in the respective cosmos of the square of the electromagnetic field, in terms of spatially natural coordinates, but in the case of $M_0$ this space is three-dimensional Euclidean space, and in the case of M it is a three-dimensional sphere, of which Euclidean space may be regarded as the stereographic projection, and the latter integral is always larger."* [S/Z95, p. 317]

The free electromagnetic field can be construed as a Hamiltonian system in infinitely many dimensions whose total energy is as just said, and the dynamical group is induced by the isometry group of temporal translations of spacetime, both in $M_0$ and in $M$.

### 12.6 Relations between the two energies

The Einstein energy $H = d/dt$ is the sum of the Minkowskian energy $H_0 = d/dx_0$ which is scale covariant (where scaling means transforming $x_j$ into $\lambda x_j$ for some constant $\lambda$ (j= 0, 1, 2, 3)) and a scale contravariant term called the *superrelativistic* energy $H_1 = -Q \, d/dx_0)Q$ where $Q$ denotes conformal inversion which on $M$ appears as the singularity-free transformation $(t,u)$ a $(\pi - t, u)$ [Seg74, p. 766 bottom right] [Seg90, p. 174], [Seg82, p. 857].

Segal suggested the following physical interpretation:

> *"Hence $H_1$ is the transform of $d/dx_0$ under conformal inversion. $H_0$ and $H_1$ correspond to effective potentials of the form $\lambda r$ and $-G/r$, where r is* the Euclidean distance." And moreover "These and other considerations are quite suggestive of the proposal that $H_1$ represents gravity, in the case of massive particles, and a generalized form of gravity, in the case of photons, so that the chronometric redshift can be regarded as gravitational in an extended sense." [Seg90, p. 174], [S/Z95, p. 309]

This hypothesis rejoins basic questions raised earlier by Penrose:

"Gravitational energy cannot be adequately defined in a local way and emerges, instead, as some kind of non local quantity. The local gravitational energy must apparently be thought of as zero, and this is consistent with Einstein's field equations." [Pen67, p. 161]

"I think we may regard the question of gravitational energy as one of the important unsolved problems in general relativity." [Pen67, p. 172]

# 13  The geometric story of light around the world

As we have seen the boundary of Minkowski spacetime as immersed in $U(2)$ is what is called *the light cone at infinity*.

This boundary is the image under conformal inversion of the light cone at the origin in $U(2)$.

A diagram essentially due to Penrose illustrates the causal immersion of Minkowski spacetime $M_0$ into the Einstein universe $M$ and also it tells what can be called the geometric story of light around the world (i.e. around $M$). [P/R86, p.294, 295], [H/E73, p. 122]

In this diagram space $S^3$ is represented as a circle $S^l$, which one may think of as a grand circle of the three-sphere, and Einstein spacetime as a cylinder whose vertical axis measures cosmic time. At cosmic time $-\pi$, infinitely long ago in local time, light was emitted at our point in space where we now stand at time *0*, the origin of Minkowski spacetime $M_0$ represented in $U(2)$ by the identity matrix $I$; this is event $i^-$ in the Penrose diagram.

This event belongs to the boundary of Minkowski spacetime as immersed in the Einstein universe. This light will forever remain on this boundary; it reaches our antipode at cosmic time *0*; this is event $i_0$ taking place at spatial infinity. Light keeps going for another half turn and will reach its departure point, the point of space where we stand, at cosmic time $+\pi$ in the infinite Minkowskian future; this is event $i^+$. It will then be ready for another turn of the merry-go-round $S^l$ representing $S^3$ in this diagram. In the diagram it would then move along a similar pattern in the next higher immersion of Minkowski spacetime in the Einstein cylinder which is not portrayed and which would extend from cosmic time $\pi$ to $3\pi$. And so on.

The three points $i^0$, $i^+$ and $\bar{i}$ are a single point in $U(2)$. Indeed they are points of the universal covering of $U(2)$ which project on the same point of $U(2)$; the last two have the same spatial coordinate and their cosmic time coordinates differ by $2\pi$. Also the first two have antipodal space coordinates and time coordinates differing by $\pi$. The last two correspond to the matrix $-I$ in U(2) which one obtains by letting $x_0$ tend to plus or minus infinity in the Cayley transform of the Hermitian matrix for $x = (x_0, 0, 0, 0)$. The point $i^0$ which is spatial infinity at time $0$ also corresponds to the matrix $-I$ which is the image of the identity matrix under conformal inversion in $U(2)$; the identity matrix being the image of the origin of $M_0$ by the Cayley transform [P/R, p. 298].

The following rendition of the Penrose diagram in question is taken from the website www.math.toronto.edu/maschler/426-termpapers/mm-GR.doc belonging to Professor Gideon Maschler, Math. Dept., University of Toronto and where more details can be found.

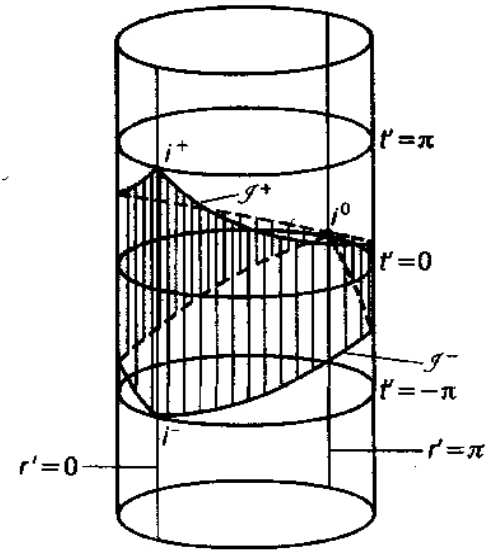

## 14 The million euro question

Could it be that the night sky were a family album of the living and the dead celestial objects, each of them being depicted a large number of times ? There would then be far fewer objects than there appears to be.

This view is now defended by a small group of Big Bang supporters [Lum03]. It would seem that this possibility arises in the context of chronometric cosmology as well.

Already in 1920, Hermann Weyl wrote [Weyl18, p. 278]:

> *"If the world is closed, spatially, it becomes possible for an observer to see several pictures of one and the same star. These depict the star at epochs separated by enormous intervals of time (during which light travels once entirely round the world)."*

In 1974, Segal wrote:

> *"In view of the apparent transparency of intergalactic space, the residual radiation should typically make many circuits of space before being ultimately absorbed by matter."*

The following excerpt from a 1995 paper by Segal and Zhou seems to imply that this theoretical possibility is in fact a prediction of CC which nevertheless has not been explicitly stated so far. Indeed in the concluding paragraph of [S/Z95] one reads:

> *"Finally, the transparency of cosmic space implies that photons in the Einstein universe EU will typically make many circuits of space (i.e. of the 3-sphere) before being absorbed or undergoing interaction. A free photon will be infinitely redshifted at the antipode of $S^3$ to its point P of emission, but on returning to P it will be in its original state, as a consequence of the periodicity of free photon wave function in EU."*

---

[4] Also with the title "Cosmological considerations on the general relativity theory" p.177-188 in "The principle of relativity", a collection of original memoirs on the special and general theory of relativity, by H.A. Lorentz, A. Einstein, H. Minkowski and H. Weyl, with notes by A. Sommerfeld, translated by W. Perrett and G.B. Jeffery; viii, 216 p. diagrs. 23 cm.. Published in London, Methuen & Co. ltd. in 1923; reprinted in 1990; also published unaltered and unabridged by Dover Publications, Inc. in 1952.

# TABLE OF CONTENTS